\documentclass[twocolumn,prl,aps,superscriptaddress,showpacs,amsmath,amssymb]{revtex4-1}
\usepackage[dvips]{graphics}
\usepackage{graphicx}
\usepackage{bm}
\usepackage{amsmath}
\usepackage{amsfonts}
\usepackage{amssymb}
\usepackage{xcolor}
\usepackage{subfigure}
\usepackage{tabularx}
\usepackage{multirow}
\usepackage{hyperref,hypcap}
\usepackage{braket}
\usepackage{commath}%
\providecommand{\U}[1]{\protect\rule{.1in}{.1in}}

\begin{document}

\title{Intrinsic Nonlinear Spin Magnetoelectricity in Centrosymmetric Magnets}

\author{Cong Xiao}
\affiliation{Department of Physics, The University of Hong Kong, Hong Kong, China}
\affiliation{HKU-UCAS Joint Institute of Theoretical and Computational Physics at Hong Kong, China}
\author{Huiying Liu}
\email{liuhuiying@pku.edu.cn}
\affiliation{Research Laboratory for Quantum Materials, Singapore University of Technology and Design, Singapore 487372, Singapore}
\author{Weikang Wu}
\email{weikang\_wu@sutd.edu.sg}
\affiliation{Research Laboratory for Quantum Materials, Singapore University of Technology and Design, Singapore 487372, Singapore}
\affiliation{Division of Physics and Applied Physics, School of Physical and Mathematical Sciences,
Nanyang Technological University, Singapore 637371, Singapore}
\author{Hui Wang}
\affiliation{Research Laboratory for Quantum Materials, Singapore University of Technology and Design, Singapore 487372, Singapore}
\author{Qian Niu}
\affiliation{School of Physics, University of Science and Technology of China, Hefei, Anhui 230026, China}
\author{Shengyuan A. Yang}
\affiliation{Research Laboratory for Quantum Materials, Singapore University of Technology and Design, Singapore 487372, Singapore}

\begin{abstract}
We propose an intrinsic nonlinear spin magnetoelectric effect
in magnetic materials, offering the potential of all-electric
control of spin degree of freedom in centrosymmetric magnets, which reside outside
of the current paradigm based on linear spin response. We reveal the band
geometric origin of this effect in the momentum and magnetization
space Berry connection polarizabilities, and clarify its symmetry
characters. As an intrinsic effect, it is determined solely by the material's band structure and represents a material characteristic. Combining our theory with first-principles calculations, we predict
sizable nonlinear spin magnetoelectricity in single-layer MnBi$_{2}$Te$_{4}$, which can be detected in experiment. Our theory paves the way for exploring rich nonlinear spintronic effects and novel device concepts based on them.
\end{abstract}
\maketitle


Electric control of spin degree of freedom is a central topic of
spintronics. Much theoretical and experimental effort \cite{Pikus1978,Aronov1989,Edelstein,Kato2004,Geller2009,Vyborny2011,Manchon2019}
has been devoted to generating spin polarization $\delta \bm s$ by an applied electric field,
characterized by a \emph{linear} spin magnetoelectric response tensor $\alpha_{ij}$ with $\delta s_i=\alpha_{ij}E_j$,
where $i$, $j$ are Cartesian indices and the Einstein summation convention is adopted.
In a magnet, the generated spin polarization may further induce torques on the magnetization and cause magnetic reversal \cite{Manchon2019}.
Importantly, since spin is even under space inversion but electric field is odd, the linear effect is constrained to systems with broken inversion symmetry~\cite{Pikus1978,Aronov1989,Edelstein,Kato2004,Geller2009,Vyborny2011,Culcer2007,Manchon2019,Garate2009,
Franz2010,Kurebayashi2014,Zelezny2014,Freimuth2014}. For the class of centrosymmetric magnets, the linear spin response is strictly forbidden in the bulk and may only occur at interfaces when forming hetero-junctions with other materials \cite{Miron2010,Miron2011,Liu2012}.


In this Letter, we unveil that sizable intrinsic nonlinear spin magnetoelectric effect can
exist in centrosymmetric magnets, thus substantially extending
the playing field of spin-charge conversion and magnetoelectricity. Since the linear response is forbidden, the leading
contribution is of the second order:
\begin{equation}
  \delta s_i=\alpha_{ij\ell} E_j E_\ell,
\end{equation}
characterized by a nonlinear response tensor $\alpha_{ij\ell}$. We show that $\alpha_{ij\ell}$ contains an \emph{intrinsic} part determined solely by the band structure. For insulators, $\alpha_{ij\ell}$ is closely connected to the electric polarization of Bloch electrons and can be expressed by an important band geometric quantity, the momentum space Berry connection polarizability (BCP).
For metals, there is an extra Fermi surface contribution, which involves a magnetization space BCP. We clarify the symmetry properties of the nonlinear intrinsic spin magnetoelectric effect. The intrinsic response has the advantage of allowing a quantitative evaluation. By combining our theory with first-principles calculations, we study the effect in single-layer MnBi$_2$Te$_4$ and find sizable result that can be detected in experiment.
Our work develops the first theory for nonlinear electric spin generation, uncovers the important roles of BCPs in spintronic effects, and opens the door to new nonlinear spintronic device concepts.
The approach here also offers a general recipe for investigating other intrinsic nonlinear response properties of
Bloch electrons.


\textcolor{blue}{\textit{Thermodynamic argument for insulators.}}
We first present a thermodynamic argument, which applies to insulating cases and captures both linear
and nonlinear spin magnetoelectric effects. It also helps to expose the role of electric polarization in the effect.

To evaluate the spin response, a conventional way is to introduce a fictitious (homogeneous) Zeeman-like field
$\bm m$ that couples to spin in the form of $-\hat{\boldsymbol{{s}}}\cdot\boldsymbol{m}$, with $\hat{\bm s}$ the spin operator.
This auxiliary field is to be distinguished from the genuine magnetization of the system and is set to zero at the end of the calculation \cite{Dong2020}.
Under the $\bm m$ field and the electric field, the electronic enthalpy of the magnetic insulator system follows the relation $d\mathcal{H}=\mathcal{-}\boldsymbol{s}\cdot d\boldsymbol{m}%
-\boldsymbol{P}\cdot d\boldsymbol{E}$, where $\boldsymbol{s}$ and $\boldsymbol{P}$ denote the spin magnetization and
the electric polarization of electrons, respectively.
According to the Maxwell relation, we have
\begin{equation}\label{max}
\frac{\partial s_{i}}{\partial E_{j}}=\frac{\partial P_{j}}{\partial m_{i}}.
\end{equation}
Therefore, the spin magnetoelectric response can be extracted from studying the electric polarization.


To include the second order spin generation, we need the polarization expanded to the first order in the electric field:
$\boldsymbol{P}=\boldsymbol{P}^{\left(  0\right)
}+\boldsymbol{P}^{\left(  1\right)  }$. The zero-field part $\bm P^{(0)}=-\int [d\bm k]\bm{\mathcal{A}}(\bm k)$ is well known \cite{KS1993,Resta1994,Xiao2010,note} (we set $e=\hbar=1$), $\boldsymbol{\mathcal{A}}\left( \boldsymbol{k}\right)
=\langle u_{n}(\bm k)|i\partial_{\boldsymbol{k}}|u_{n%
}(\bm k)\rangle$ is the intraband Berry connection for a band eigenstate $|u_{n}(\bm k)\rangle$, $[d\bm k]$ is a short-hand notation for $\sum_n d\bm k/(2\pi)^d$ with $d$ being the dimension of the system, and the summation is over all occupied bands.
Here and hereafter, for simple notations, we drop the band index $n$ wherever appropriate.
$\bm P^{(0)}$ gives the linear spin magnetoelectric response with $\delta\bm s=\partial_{\bm m}P^{(0)}_j E_j$, reproducing the result in previous works \cite{Yuriy2017,Xiao2021}.


On the other hand, the second order spin response is contained in $\bm P^{(1)}$, the polarization linear in ${\bm E}$ field~\cite{Sipe1995,Nunes2001,Vanderbilt2002}.
It can be expressed as $\bm P^{(1)}=-\int [d\bm k] \bm a(\bm k)$~\cite{Gao2014} in terms of the field-induced Berry connection  $a_i=G_{ij} E_j$, where
\begin{equation}\label{k-BCP}
  G_{ij}=2 \text{Re}\sum_{n'\neq n}\frac{(v_i)_{nn'}(v_j)_{n'n}}{(\varepsilon_{n}-\varepsilon_{n'})^3}
\end{equation}
is known as the BCP for the state $|u_{n}(\bm k)\rangle$~\cite{Gao2014}, $(v_i)_{nn'}$
is the interband velocity matrix element, and $\varepsilon_{n}$ is the unperturbed energy for $|u_{n}(\bm k)\rangle$. Combining this result with Eq.~(\ref{max}), we immediately find the second order spin polarization
\begin{equation}\label{nl}
  \delta\bm s=\partial_{\bm m}\Big[ \frac{1}{2}\bm E\cdot\bm P^{(1)}\Big]
\end{equation}
with the nonlinear response tensor
\begin{equation}\label{insu}
  \alpha_{ij\ell}=-\frac{1}{2}\partial_{m_i}\int [d\bm k]G_{j\ell}(\bm k)
\end{equation}
expressed nicely in terms of the momentum space BCP of occupied states.
Recent studies have highlighted the role of this BCP in nonlinear transport phenomena~\cite{Gao2014,Gao2017,Gao2018,Xiao2021adiabatic,Lai2021,Wang2021,Liu2021,Liu2022}, whereas our result here unveils its significance in magnetoelectric effects.

The thermodynamic argument reveals the important role of electric polarization in the spin magnetoelectricity. Particularly, the term in the bracket of Eq.~(\ref{nl}) is just the material-dependent part of the electric energy density in a dielectric that is of $E^2$ order.
For centrosymmetric systems, the zero-field polarization $\bm P^{(0)}$ vanishes, so the argument confirms that the linear spin response must also vanish. Meanwhile, $\bm P^{(1)}$ can be nonzero regardless of the inversion symmetry, so that the nonlinear spin response would become dominant
in centrosymmetric materials.

It is also important to note that the argument above applies only to insulators. For metals, the electric polarization ceases to be well defined, hence we need a more general approach to the problem. As we shall see, in a magnetic metal, there will be additional nonlinear contributions from the Fermi surface.

\textcolor{blue}{\textit{Intrinsic nonlinear spin magnetoelectricity.}}
To establish a general result which is applicable also to metallic cases, we develop a semiclassical theory for Bloch
electrons in the nonlinear response regime. In Refs. \cite{Gao2014,Gao2015}, Gao \emph{et al.} extended the semiclassical theory to second order accuracy.
Nonetheless, the formulation there is focusing on the charge degree of freedom, but does not explicitly handle spin. Here, we add this missing piece. As is detailed in the Supplemental Material \cite{supp}, within the extended semiclassical
framework, we derive the following spin expectation value corrected to second order of $E$ field for an electron wave packet centered at $|u_{n}(\bm k)\rangle$:
\begin{equation}\label{sn}
  \bm s_{n }(\bm k)=-\partial_{\bm m} \tilde{\varepsilon}_n+ \bm{{\Omega}}_{\bm m\bm k}\cdot \bm E.
\end{equation}
Here, $\tilde{\varepsilon}_n=\varepsilon_n-(1/2) G_{ij}E_i E_j$ is the field-corrected band energy, and $(\bm{{\Omega}}_{\bm m\bm k})_{ij}=\partial_{m_i}
(\mathcal{A}_j+a_j)-\partial_{k_j}(\mathfrak{A}_i+\mathfrak{a}_i)$ is the field-corrected Berry curvature in the hybrid $k$-$m$ space.
The definitions of the Berry connections $\mathfrak{\bm A}$ and $\mathfrak{\bm a}$ are analogous to their counterparts $\mathcal{\bm A}$ and ${\bm a}$ in $k$ space. Specifically, $\mathfrak{A}_i=\langle u_n(\bm k)|i\partial_{m_i}|u_n(\bm k)\rangle$, and $\mathfrak{a}_i=\mathfrak{G}_{ij}E_j$ can be expressed using a $m$-space BCP
\begin{equation}\label{m-BCP}
  \mathfrak{G}_{ij}=-2 \text{Re}\sum_{n'\neq n}\frac{(s_i)_{nn'}(v_j)_{n'n}}{(\varepsilon_{n}-\varepsilon_{n'})^3},
\end{equation}
where the numerator involves the interband matrix elements of spin and velocity operators. Like $G_{ij}$, $\mathfrak{G}_{ij}$ is gauge invariant, so it is also an intrinsic band geometric property.

\renewcommand{\arraystretch}{1.42} \begin{table*}[ptb]
\caption{Constraints on the intrinsic nonlinear spin magnetoelectric response
tensor elements from magnetic point group
symmetries. \textquotedblleft$\checkmark$\textquotedblright%
\ (\textquotedblleft$\times$\textquotedblright) means that the element is
symmetry allowed (forbidden). Here, we choose to symmetrize the second and the third tensor indices, by defining $\alpha_{i(xy)}\equiv\frac{1}{2}(\alpha
_{ixy}+\alpha_{iyx})$, and we omit the superscript `int' in the table. Symmetry operations $\mathcal{T}$, $\mathcal{PT}$,
$C_{3}\mathcal{T}$, and $S_{6}\mathcal{T}$ forbid all the
elements here, hence are not listed.}%
\label{tab:sym}
\begin{centering}
\begin{tabular}{p{0.06\linewidth}p{0.04\linewidth}<{\centering}p{0.04\linewidth}<{\centering}
p{0.05\linewidth}<{\centering}p{0.05\linewidth}<{\centering}p{0.05\linewidth}<{\centering}
p{0.05\linewidth}<{\centering}p{0.04\linewidth}<{\centering}p{0.04\linewidth}<{\centering}
p{0.045\linewidth}<{\centering}p{0.05\linewidth}<{\centering}p{0.05\linewidth}<{\centering}
p{0.05\linewidth}<{\centering}p{0.045\linewidth}<{\centering}p{0.05\linewidth}<{\centering}
p{0.045\linewidth}<{\centering}p{0.045\linewidth}<{\centering}p{0.045\linewidth}<{\centering}}
\hline \hline
& \multirow{2}{*}{$\mathcal{P}$} & \multirow{2}{*}{$C_{2}^{z}$} & $C_{3}^{z}$, & $C_{4,6}^{z}$, & $C_{2,4,6}^{x}$, & $C_{3}^{x}$, & \multirow{2}{*}{$\sigma_{z}$} & \multirow{2}{*}{$\sigma_{x}$} & \multirow{2}{*}{$C_{2}^{z}\mathcal{T}$} & $C_{4}^{z}\mathcal{T}$, & \multirow{2}{*}{$C_{6}^{z}\mathcal{T}$} & \multirow{2}{*}{$C_{2}^{x}\mathcal{T}$} & $C_{4}^{x}\mathcal{T}$, & \multirow{2}{*}{$C_{6}^{x}\mathcal{T}$} & \multirow{2}{*}{$\sigma_{z}\mathcal{T}$} & \multirow{2}{*}{$\sigma_{x}\mathcal{T}$}\tabularnewline[-1.2ex]
& & &$S_{6}^{z}$  & $S_{4}^{z}$ & $S_{4}^{x}$ & $S_{6}^{x}$ & & & &$S_{4}^{z}\mathcal{T}$  & & &$S_{4}^{x}\mathcal{T}$ & & & \tabularnewline\hline
\ $\alpha_{xxx}$ & $\checkmark$ & $\times$ & $-\alpha_{xyy}$ & $\times$ & $\checkmark$ & $\checkmark$ & $\times$ & $\checkmark$ & $\checkmark$ & $\times$ & $-\alpha_{xyy}$ & $\times$ & $\times$ & $\times$ & $\checkmark$ & $\times$\tabularnewline
\ $\alpha_{x(xy)}$ & $\checkmark$ & $\times$ & $\alpha_{yxx}$ & $\times$ & $\times$ & $\times$ & $\times$ & $\times$ & $\checkmark$ & $\times$ & $\alpha_{yxx}$ & $\checkmark$ & $\times$ & $\times$ & $\checkmark$ & $\checkmark$\tabularnewline
\ $\alpha_{xyy}$ & $\checkmark$ & $\times$ & $\checkmark$ & $\times$ & $\checkmark$ & $\checkmark$ & $\times$ & $\checkmark$ & $\checkmark$ & $\times$ & $\checkmark$ & $\times$ & $\checkmark$ & $\times$ & $\checkmark$ & $\times$\tabularnewline
\ $\alpha_{yxx}$ & $\checkmark$ & $\times$ & $\checkmark$ & $\times$ & $\times$ & $\times$ & $\times$ & $\times$ & $\checkmark$ & $\times$ & $\checkmark$ & $\checkmark$ & $\times$ & $\times$ & $\checkmark$ & $\checkmark$\tabularnewline
\ $\alpha_{y(xy)}$ & $\checkmark$ & $\times$ & $\alpha_{xyy}$ & $\times$ & $\checkmark$ & $\checkmark$ & $\times$ & $\checkmark$ & $\checkmark$ & $\times$ & $\alpha_{xyy}$ & $\times$ & $\checkmark$ & $\times$ & $\checkmark$ & $\times$\tabularnewline
\ $\alpha_{yyy}$ & $\checkmark$ & $\times$ & $-\alpha_{yxx}$ & $\times$ & $\times$ & $\checkmark$ & $\times$ & $\times$ & $\checkmark$ & $\times$ & $-\alpha_{yxx}$ & $\checkmark$ & $\times$ & $\checkmark$ & $\checkmark$ & $\checkmark$\tabularnewline
\ $\alpha_{zxx}$ & $\checkmark$ & $\checkmark$ & $\checkmark$ & $\checkmark$ & $\times$ & $\times$ & $\checkmark$ & $\times$ & $\times$ & $\checkmark$ & $\times$ & $\checkmark$ & $\times$ & $\times$ & $\times$ & $\checkmark$\tabularnewline
\ $\alpha_{z(xy)}$ & $\checkmark$ & $\checkmark$ & $\times$ & $\times$ & $\checkmark$ & $\checkmark$ & $\checkmark$ & $\checkmark$ & $\times$ & $\checkmark$ & $\times$ & $\times$ & $\checkmark$ & $\times$ & $\times$ & $\times$\tabularnewline
\ $\alpha_{zyy}$ & $\checkmark$ & $\checkmark$ & $\alpha_{zxx}$ & $\alpha_{zxx}$ & $\times$ & $\checkmark$ & $\checkmark$ & $\times$ & $\times$ & $-\alpha_{zxx}$ & $\times$ & $\checkmark$ & $\times$ & $\checkmark$ & $\times$ & $\checkmark$\tabularnewline
\hline \hline
\end{tabular}
\par\end{centering}
\end{table*}

With the spin polarization for each state, the total spin polarization in the system can be obtained as
\begin{equation}\label{ss}
  \bm s=\int [d\bm k] \bm s_n(\bm k) f_n(\bm k),
\end{equation}
where $f$ is the electron distribution function. The second-order spin response is obtained by inserting the expression in (\ref{sn}) and retaining terms that are of $E^2$ order. Here, we are particularly interested in the intrinsic contribution that involves only the equilibrium Fermi distribution function $f_0({\varepsilon}_n)$ of the unperturbed band structure. The intrinsic nonlinear response tensor is obtained as
\begin{equation}\begin{split}\label{key}
  \alpha_{ij\ell}^\text{int}=-\frac{1}{2}\partial_{m_i}\int [d\bm k]G_{j\ell}f_0
  -\int [d\bm k] (s_i G_{j\ell}+v_j \mathfrak{G}_{i\ell})f_0',
  \end{split}
\end{equation}
where $s_i$ ($v_j$) are the intraband spin (velocity) matrix elements for the state $|u_n(\bm k)\rangle$.

Equation (\ref{key}) is the key result of this work. First, it applies to both insulators and metals. Compared to (\ref{insu}), Eq.~(\ref{key}) contains an additional Fermi surface term (the second term). For the case of an insulator, the Fermi surface term vanishes, and the result recovers (\ref{insu}), confirming the consistency between the two approaches.
Second, $\alpha_{ij\ell}^\text{int}$ is suppressed by time reversal symmetry, and is nonzero only for magnetic systems, as can be verified directly from Eq.~(\ref{key}).
Third, as an intrinsic contribution, Eq.~(\ref{key}) is a genuine material property, determined solely by the material's band structure.
{We mention that the $m$-derivative in the first term of (\ref{key}) can be done straightforwardly to obtain an expression involving only the spin and velocity matrix elements [see Eq.~(S6) in \cite{supp}] of the band structure. Hence, the response can be readily evaluated in first-principles calculations.
}

\textcolor{blue}{\textit{Symmetry property.}}
As we have discussed, the intrinsic nonlinear response is not suppressed by the inversion symmetry. Meanwhile, other crystalline symmetries also put constraints on the form of $\alpha_{ij\ell}^\text{int}$, which we analyze here.

Since spin is a time reversal ($\mathcal{T}$) odd pseudovector and the electric field is a $\mathcal{T}$ even vector, $\alpha_{ij\ell}^\text{int}$ transforms as a third-rank $\mathcal{T}$ odd pseudotensor, which respects
\begin{equation}
\alpha^\text{int}_{i^{\prime}j^{\prime}\ell^{\prime}}=\eta_T\text{det}(O)O_{i^{\prime}%
i}O_{j^{\prime}j}O_{\ell^{\prime}\ell}\alpha^\text{int}_{ij\ell}. \label{eq:constraints}%
\end{equation}
{Here $O$ is a point group operation, and the factor $\eta_T=\pm$ is connected with the character of $\alpha_{ij\ell}^\text{int}$ being $\mathcal{T}$ odd: $\eta_T=-1$ for primed operations, i.e., the magnetic symmetry operations of the form $R\mathcal{T}$ with $R$ a spatial operation;
and $\eta_T=+1$ for nonprimed operations.}

Assuming the applied electric field is in the $x$-$y$ plane, the constraints from different magnetic point group symmetries are listed in
Table~\ref{tab:sym}. This offers useful guidance for analyzing the nonlinear spin response for a particular material. For example, consider a ferromagnet which preserves the inversion and a horizontal mirror $\sigma_z$, with its magnetization along the $z$ direction. Then, Table~\ref{tab:sym} tells us that for an applied in-plane electric field, the generated nonlinear spin polarization must be out-of-plane, along the magnetization direction. For the purpose to induce magnetic reversal, one may want the induced spin polarization to have a component normal to the magnetization, such that it can generate a torque, which means that the desired material should not have a horizontal mirror plane.

\begin{figure}[ptb]
\centering
\includegraphics[width=0.5\textwidth]{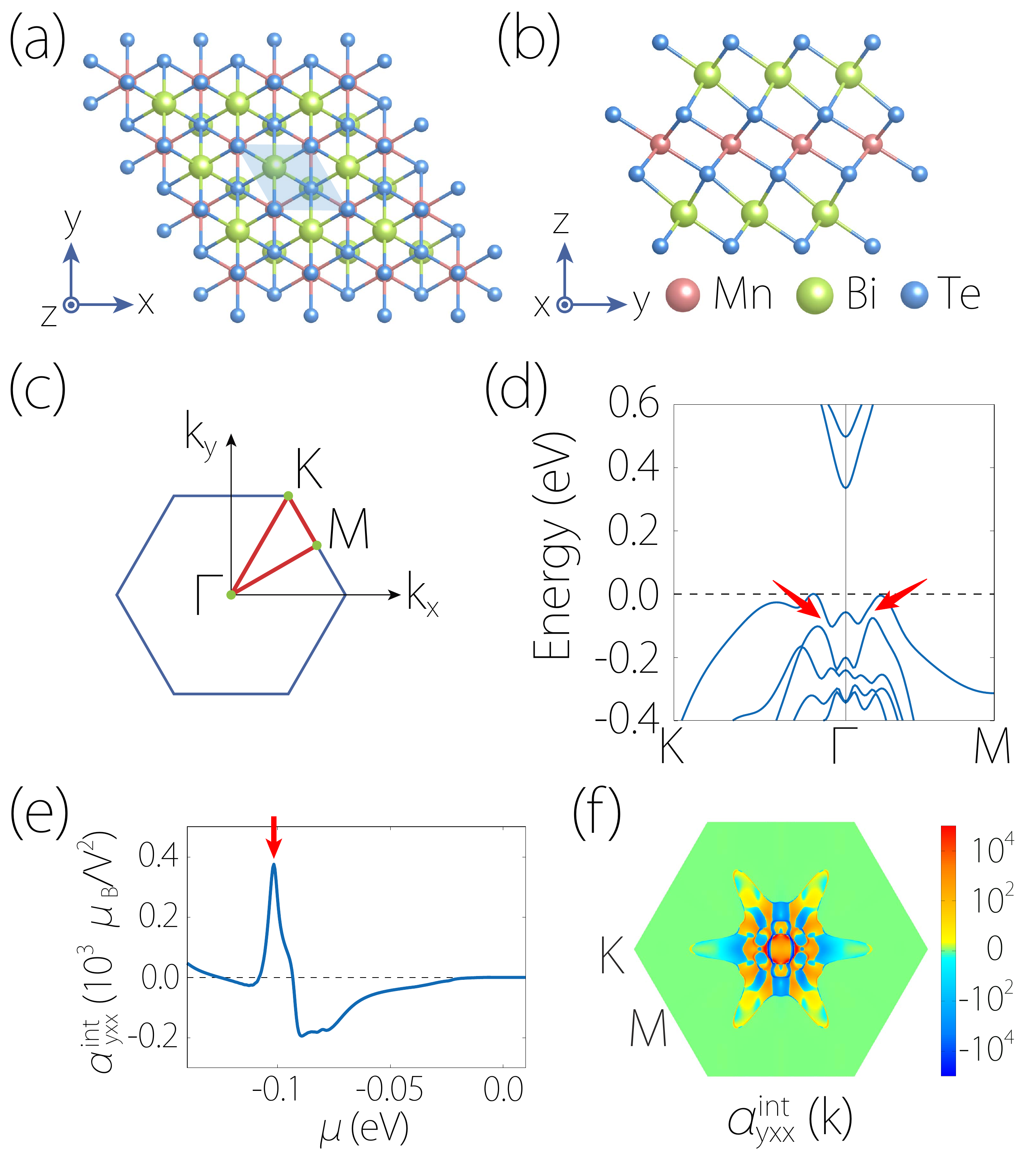} \caption{(a) Top and
(b) side views of the lattice structure of single-layer MnBi$_{2}$Te$_{4}$. (c) shows
the Brillouin zone. (d) Calculated band structure (spin-orbit coupling is included). (e) Calculated nonlinear response tensor element $\alpha_{yxx}^\text{int}$ versus
the chemical potential $\mu$. (f) Distribution of $\alpha_{yxx}^\text{int} (\bm{k})$, i.e., the integrand of Eq.~(\ref{key}), in the momentum space for chemical potential at
$-0.1$~eV [marked by the red arrow in (e)]. The unit is $\mu_{B}/\text{V}^{2}$ per unit cell. In the
calculation, the temperature is set to 8~K.}%
\label{fig:MBT_ML}%
\end{figure}

\textcolor{blue}{\textit{A material example.}}
We demonstrate the implementation of our theory in first-principles calculations to study a concrete material.
Guided by the symmetry constraints in Table~\ref{tab:sym}, we consider the effect in single-layer MnBi$_{2}$Te$_{4}$.
MnBi$_{2}$Te$_{4}$ in its bulk and two dimensional few-layer forms has attracted considerable research interest recently, because
it provides a platform for realizing various types of topological states
\cite{Gong2019,Chulkov2019,Deng2020,Zhang2019,Xu2019,Otrokov2019,Zeugner2019,Yan2019,Lee2019,Cui2019,Liu2020}.
Our focus here is on its single layer, which has been
successfully fabricated in experiment, either by exfoliation from the bulk or by molecular
beam epitaxy growth \cite{Deng2020,Liu2020,Gong2019}. Its crystal structure is shown in Fig.~\ref{fig:MBT_ML}(a)
and~\ref{fig:MBT_ML}(b), characterized by a hexagonal lattice with the space
group $P\bar{3}m1$ (No.~164) and the point group $D_{3d}$. It consists of seven atomic layers, stacked in the sequence of
Te-Bi-Te-Mn-Te-Bi-Te. Previous works have established that the ground state of single-layer MnBi$_{2}$Te$_{4}$ is a
topologically trivial ferromagnetic semiconductor with out-of-plane magnetization, and the Curie temperature is about 12 K \cite{Chulkov2019}.
The ground-state magnetic configuration
possesses a magnetic point group of $\bar{3}m^{\prime}$. Importantly, the inversion
symmetry is preserved, which forbids the linear spin magnetoelectric response. According to Table~\ref{tab:sym}, the
symmetries $C_{3}^{z}$, $C_{2}^{x}\mathcal{T}$, and
$\sigma_{x}\mathcal{T}$ further enforce the following relations among
elements of the nonlinear response tensor: $\alpha_{yxx}^\text{int}=\alpha
_{xxy}^\text{int}=\alpha_{xyx}^\text{int}=-\alpha_{yyy}^\text{int}$ and $\alpha_{zxx}^\text{int}=\alpha_{zyy}^\text{int}$.


It follows that the nonlinear spin magnetoelectric response of single-layer MnBi$_{2}$Te$_{4}$ is
specified by only two independent elements $\alpha_{yxx}^\text{int}$ and $\alpha_{zxx}^\text{int}$.
To see this more clearly, we assume the electric field is along an in-plane direction that makes an angle $\theta$ from the $x$ axis,
 i.e., $\boldsymbol{E}=E(\cos\theta,\sin\theta,0)$. The induced out-of-plane
nonlinear spin polarization takes the form of
$\delta s_{z}=\alpha_{zxx}^\text{int}E^{2}$, which is
independent of the field direction. Meanwhile, the induced in-plane spin polarization can
be expressed as \cite{supp}
\begin{equation}
(\delta s_{\parallel},\delta s_{\perp})=\alpha_{yxx}^\text{int}(\sin3\theta,\cos
3\theta)E^{2}, \label{eq:yxx}%
\end{equation}
where $\delta s_{\parallel}$ and $\delta s_{\perp}$ denote the components
parallel and perpendicular to the electric field, respectively. It is
interesting to note that the in-plane nonlinear response is characterized by a single
element $\alpha_{yxx}^\text{int}$, and exhibits an angular dependence with $2\pi/3$ periodicity. Since we are more interested in the spin polarization that is normal to the magnetization direction, we shall focus on  $\alpha_{yxx}^\text{int}$ in the following discussion.

Next, we evaluate the intrinsic nonlinear spin magnetoelectric response tensor by combining our theory with first-principles calculations
(the calculation details are presented in \cite{supp}). Figure~\ref{fig:MBT_ML}(d) shows the calculated
band structure of single-layer MnBi$_{2}$Te$_{4}$.
The system is a ferromagnetic semiconductor with an indirect gap of 337~meV,
which agrees with previous result~\cite{Chulkov2019}.
We have computed all the relevant tensor elements for $\alpha_{ij\ell}^\text{int}$ according to Eq.~(\ref{key}).
The results comply with the symmetry constraints discussed above. As mentioned, we
focus on the in-plane spin generation. The obtained $\alpha_{yxx}^\text{int}$ as a function of
the chemical potential is plotted in Fig.~\ref{fig:MBT_ML}(e).
Within the large band gap, the value of $\alpha_{yxx}^\text{int}$ is small but nonzero. It is $\sim-0.036$~$\mu_{\text{B}}%
$/V$^{2}$, with $\mu_{\text{B}}$ as the Bohr magneton.
The response is greatly enhanced by hole doping, especially when the chemical potential is
shifted to band near degeneracy regions in the valence bands.
Because Eqs.~(\ref{k-BCP}) and (\ref{m-BCP}) show that the BCPs are generally large around band near degeneracies,
it follows that $\alpha_{ij\ell}^\text{int}$, involving integrals of BCPs,
must also be peaked when the chemical potential is aligned in such regions.
Particularly, in Fig.~\ref{fig:MBT_ML}(e), a peak of $375$~$\mu_{\text{B}}$/V$^{2}$ is observed
around $-0.1$~eV, which can be attributed to the small gap regions marked by red arrows in Fig.~\ref{fig:MBT_ML}(d). At the peak, we plot the
$k$-resolved contribution to $\alpha_{yxx}^\text{int}$, i.e., the integrand in Eq.~(\ref{key}), in
Fig.~\ref{fig:MBT_ML}(f). The distribution shows an even function with
respect to both the $x$ and $y$ axis, and is peaked around the small-gap regions.

Consider the hole doped case with the chemical potential $\sim -0.1$~eV and
a moderate applied electric field of 1~kV/cm which is readily achievable in experiment.
The induced in-plane spin magnetization in single-layer MnBi$_{2}$Te$_{4}$ is $\sim 0.4\times10^{-5}$
$\mu_{\text{B}}$/nm$^{2}$, which is comparable to the reported linear spin generation
in typical noncentrosymmetric systems~\cite{Mertig2016,Rossi2017,Manchon2019}.
Hence, it should be detectable in experiment, e.g., by magneto-optical Kerr spectroscopy,
and it can produce considerable spin-orbit torque effects.
The effect can be further enhanced by more than an order of magnitude at higher doping levels
$\sim -0.16$~eV (shown in \cite{supp}). In practice, the doping can be readily controlled for 2D materials by gating.


\textcolor{blue}{\textit{Discussion.}}
We have presented the first theory of nonlinear spin magnetoelectric effect, which is the leading response in centrosymmetric magnets. The focus here is on the intrinsic contribution, which can be quantitatively evaluated for each material. For insulators, it captures the total response, whereas for metals, there are additional extrinsic contributions from the nonequilibrium distribution at the Fermi surface. The extrinsic contributions are in principle also contained in Eq.~(\ref{ss}), and can be extracted by solving the distribution function, e.g., from the Boltzmann equation. They are connected with carrier scattering and will involve the relaxation time parameter. A systematic study of the extrinsic effect is an interesting topic to explore in future works. In practice, the intrinsic and extrinsic parts can be separated by their different scaling with the relaxation time and distinct symmetry constraints, analogous to cases in nonlinear charge transport \cite{Lai2021,Wang2021,Liu2021,Fu2015,Kang2019,Du2019}.

We have demonstrated the implementation of our theory with first-principles calculations. This will
guide the experimental study and facilitate the search for nonlinear spintronic material platforms.
The effect should exist in conventional ferromagnets like fcc Ni and Co, which preserve the inversion symmetry.
We also expect the recently fabricated 2D centrosymmetric magnets, such as 1T-MnSe$_{2}$ \cite{Kawakami2018},
CrI$_{3}$ \cite{Xu2017,Xu2018}, and 1T-VSe$_{2}$ \cite{Bonilla2018}, would be good candidates, due to their great tunability.

Finally, we note that intrinsic second order responses of other observables that correspond to
local operators, such as charge current \cite{Gao2014},
spin current, and pseudospin, admit a similar formulation as the theory developed here (see the
Supplemental Material \cite{supp}).
Therefore, our finding not only serves as a building block for the emerging field of nonlinear spintronics,
but also forms the basis for exploring rich intrinsic nonlinear response properties of Bloch electrons.


\begin{thebibliography}{99}
\bibitem{note}Here we assume the system has no nontrivial Chern number.
\end{thebibliography}
\end{document}